\documentclass[a4paper,11pt]{article}
\usepackage{jinstpub}
\usepackage{lineno}
\usepackage{placeins}
\title{\boldmath A borehole muon detector with SiPM-on-tile technology}
\author[a,1]{Miguel Arratia\note{Corresponding author.}}
\author[a]{Jiajun Huang}
\author[a]{Sean Preins}
\author[d]{Sebastian Ritter}
\author[b,c]{Christian P. Romero}
\author[b,c]{Sebastian Tapia}
\affiliation[a]{Department of Physics and Astronomy, University of California, Riverside, CA 92521, USA}
 \affiliation[b]{Departamento de Física, Universidad Técnica Federico Santa María, Valparaíso, Chile}
 \affiliation[c]{Centro Cientifico Tecnólogico de Valparaíso (CCTVal), Universidad Técnica Federico Santa María, Valparaíso, Chile }
\affiliation[d]{Institut f\"ur Physik, Johannes Gutenberg-Universit\"at Mainz, 55128 Mainz, Germany}

\emailAdd{miguel.arratia@ucr.edu}

\abstract{We developed a compact and rugged muon detector designed for deployment in boreholes. The detector uses a SiPM-on-tile approach in which silicon photomultipliers are directly coupled to scintillator tiles, thereby eliminating the need for wavelength-shifting fibers and long scintillator bars. The modular design is based on a 64-channel unit, 140~cm in length and 80~mm in diameter, composed of $5 \times 5$~cm$^{2}$ scintillator tiles coupled to SiPMs, powered and read out using off-the-shelf electronics. The detector has an average muon detection efficiency above 95\% and acceptance over 5$^\circ$--60$^\circ$ in zenith and 0$^\circ$--360$^\circ$ in azimuth. Simulations indicate that reconstruction combining hit positions and energy deposits achieves a zenith resolution of 1.5--4.0$^\circ$ across most of the zenith range. This work demonstrates a compact, rugged, and cost-effective borehole muon detector based on the SiPM-on-tile approach, offering a new alternative for muon tomography.}

\begin{document}
\maketitle
\setcounter{tocdepth}{2}

\tableofcontents
\bigskip

\flushbottom

\section{Introduction}
Muon tomography is a well-established technique for reconstructing the three-dimensional density of objects~\cite{Tanaka2023Muography,IAEA2022MuonImaging}. Since detectors can only probe volumes above their position, several designs have been developed for deployment in boreholes~\cite{Levy1985BoreholeMuon,Schwitters2004MuonTomography,fermi2007university,SchreinerIII:2016ezv,BASSET2006298,Klinger:2015gva,Gluyas2018CO2Muon,Yaish2017GasMuon,Harel2019Lingacom,Bonneville2017MuonDetector,Flygare2018MuonBorehole,Olah2018MWPCBorehole,Hamar2022Portable,HiroshiSuenaga2018,Schouten:2018pqz,Schouten2024Helical,Holma2020_US11892463B2,Schouten_drift,Cimmino2021BoreholeDetector,Ambrosino2022Borehole,Saracino2023Borehole,Gauvreau2024B2,Yamazaki2022,Yao:2023oyn,Li2024Omnidirectional,Zhao_2024,Yin2024Borehole}. Existing systems broadly fall into scintillator- and gas-based detectors, as summarized in Table~\ref{tab:borehole_detectors}. 

Among these, plastic scintillator designs are the most mature. However, three-dimensional hit measurement requires either multiple layers, which increase the detector diameter and limit compatibility with smaller boreholes, or picosecond-level timing electronics when long bars or fibers are read out from both ends to infer the longitudinal coordinate of hits.

We address these limitations with a compact design based on scintillator tiles individually read out by SiPMs, which provide three-dimensional hit positions while reducing detector diameter and easing timing requirements. This concept follows the SiPM-on-tile technology developed for collider-physics calorimetry~\cite{Sefkow:2015hna,CALICE:2022uwn,CMS:2022jvd,Belloni_2021}, where it has largely replaced fiber-based readout.

\begin{table}[ht]
\centering
\caption{Summary of borehole muon detector designs, ordered from most recent to earliest. For simplicity, the term ``bar'' refers to scintillators with one elongated dimension, although cross-sectional shapes may vary and include trapezoids, rectangles, wedges, and other geometries. SiPM sizes are reported where available. Acronyms used include SiPM (silicon photomultiplier), PMT (photomultiplier tube), WLS (wavelength shifting), CsI (cesium iodide), TPC (time projection chamber), MPGD (micropattern gaseous detector), and MWPC (multi-wire proportional chamber).}
\label{tab:borehole_detectors}
\begin{tabular}{|l|p{5.5cm}|c|r|l|}
\hline
Year \& Ref. & Detector technology & Ø (mm) & Ch./Length & SiPM size \\
\hline
2025 -- This work & Tile scintillator + SiPM & 80  & 64/0.9 m  & 1.3 mm \\
\hline
2024 -- Yin et al.~\cite{Yin2024Borehole} & Arc \& bar scintillator + SiPM & 100 & 176/1.0 m  & 6 mm \\
\hline
2024 -- Zhao et al.~\cite{Zhao_2024} & Cylindrical scintillator + WLS fiber + SiPM & 143 & 64/0.23 m & n.a\\
\hline 
2024 -- Li et al.~\cite{Li2024Omnidirectional} & Annular \& bar scintillator + WLS fiber + SiPM & 250 & 148/1.5 m  & 6 mm\\  
\hline 
2023 -- Yao et al.~\cite{Yao:2023oyn} & CsI + SiPM & 100 & 34/1.0 m  & 6 mm \\
\hline
2022 -- Yamazaki et al.~\cite{Yamazaki2022} & Bar scintillator + WLS fiber + PMT & 90 & 32/1.0 m  & n.a \\
\hline
2022 -- Gauvreau et al.~\cite{Gauvreau2024B2} & TPC + MPGD & $<$90 & n.s & n.a \\
\hline
2021 -- Cimmino et al.~\cite{Cimmino2021BoreholeDetector,Ambrosino2022Borehole,Saracino2023Borehole} & Arc \& bar scintillator + SiPM & 240 & 384/1.0 m & 3 mm \\
\hline
2021 -- Schouten et al.~\cite{Schouten_drift} & Drift tubes & $<$89 & n.s & n.a \\
\hline
2020 -- Holma et al.~\cite{Holma2020_US11892463B2} & Bar scintillator + SiPM & $<$100 & n.s & n.s \\
\hline
2018 -- Schouten et al.~\cite{Schouten:2018pqz,Schouten2024Helical} & Helix scintillator + bar scintillator + WLS fiber + PMT & $<$89 & 128/1.0 m    & n.a \\
\hline 
2018 -- Suenaga et al.~\cite{HiroshiSuenaga2018} & Scintillator bars + WLS fiber + PMT & 114 &124/2.4 m & n.a \\ 
\hline
2018 -- Olah et al.~\cite{Olah2018MWPCBorehole,Hamar2022Portable} & MWPC & 100 & 64/0.2 m  & n.a \\
\hline
2017 -- Bonneville et al.~\cite{Bonneville2017MuonDetector,Flygare2018MuonBorehole} & Bar scintillator + WLS fiber + SiPM  & 150 & 90/0.7 m & 3 mm \\
\hline
2017 -- Yaish et al.~\cite{Yaish2017GasMuon,Harel2019Lingacom} & MPGD & n.s & n.s & n.a \\ 
\hline 
2015 -- Klinger et al.~\cite{Klinger:2015gva,Gluyas2018CO2Muon} & Bar scintillator + PMT & 230 & 14/1.0 m & n.a\\ 
\hline 
2006 -- Basset et al.~\cite{BASSET2006298} & Bar and fiber scintillator + PMT & 140 & 384/2.2 m & n.a \\
\hline
2004 -- Schwitters et al.~\cite{Schwitters2004MuonTomography,fermi2007university,SchreinerIII:2016ezv} & Helix scintillator + WLS fiber + PMT, and gas Cherenkov & 1500 & 448/4.2m & -- \\ 
\hline 
1981 -- Levy et al.~\cite{Levy1985BoreholeMuon} & MWPC  & n.s. & n.s. & n.a \\
\hline

\hline

\end{tabular}
\end{table}


\FloatBarrier
\section{Borehole detector design}
\label{sec:design}
The detector design is modular, with each unit (``SiPM board'') consisting of a printed circuit board equipped with eight SiPMs (1.3 mm, Hamamatsu S14160-1315PS), each air-coupled to a 4 mm thick EJ-212 scintillator tile ($5 \times 5$ cm$^{2}$) featuring a central spherical dimple of radius \(3.8\,\mathrm{mm}\) and depth \(1.6\,\mathrm{mm}\), following Ref.~\cite{Belloni_2021}. The tiles are wrapped in enhanced specular reflector (ESR) foil to maximize light yield and covered with polyimide film for protection. The analog signals from each unit are routed to the readout electronics via a flat cable. 

Eight SiPM boards form a four-sided detector with 64 channels in total, measuring 90~cm in length and 80~mm in diameter. The detector skeleton, to which the SiPM boards are mounted, is fabricated by 3D printing. To improve polar-angle reconstruction, two of the four sides are shifted longitudinally by a quarter of a cell; to improve azimuthal-angle resolution, the two detector halves are rotated by 45$^\circ$ relative to each other.

For our prototype we used the CAEN A5202, which provides the power supply and trigger/readout for 64 SiPMs, as well as two temperature sensor inputs. The CAEN unit and auxiliary sensors (e.g., temperature and humidity monitors) are controlled by a Raspberry~Pi. In the borehole configuration, power is supplied externally via umbilical cables. The detector is encased in a waterproof carbon-fiber housing.

\begin{figure}[h!]
\centering
\includegraphics[width=0.72\linewidth]{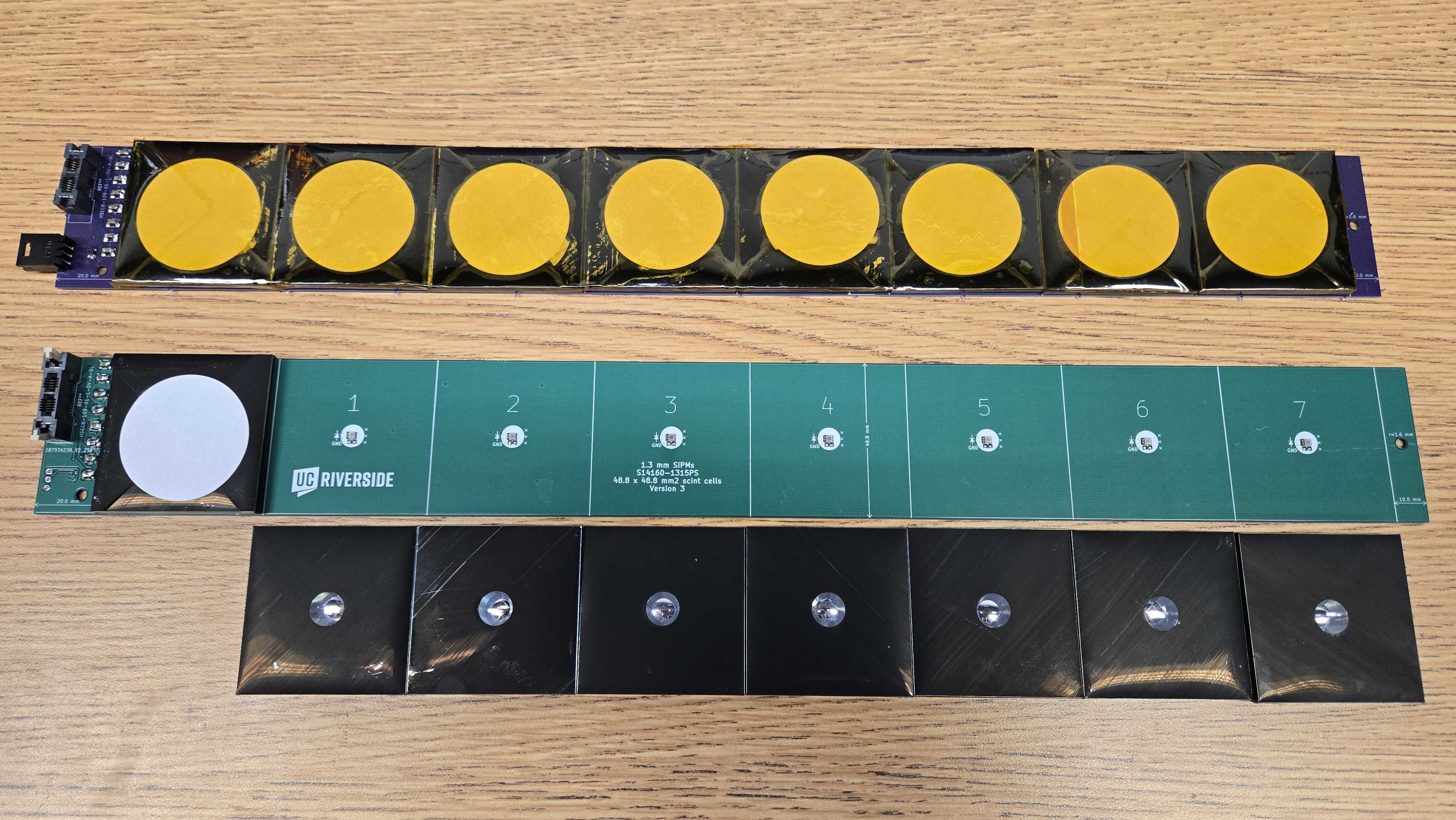}
\includegraphics[width=0.26\linewidth]{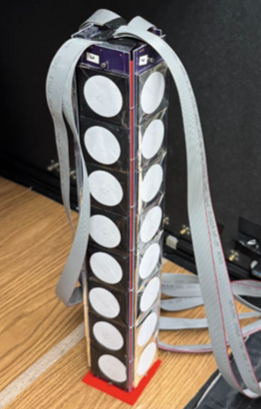}\\
\includegraphics[width=1.0\linewidth]{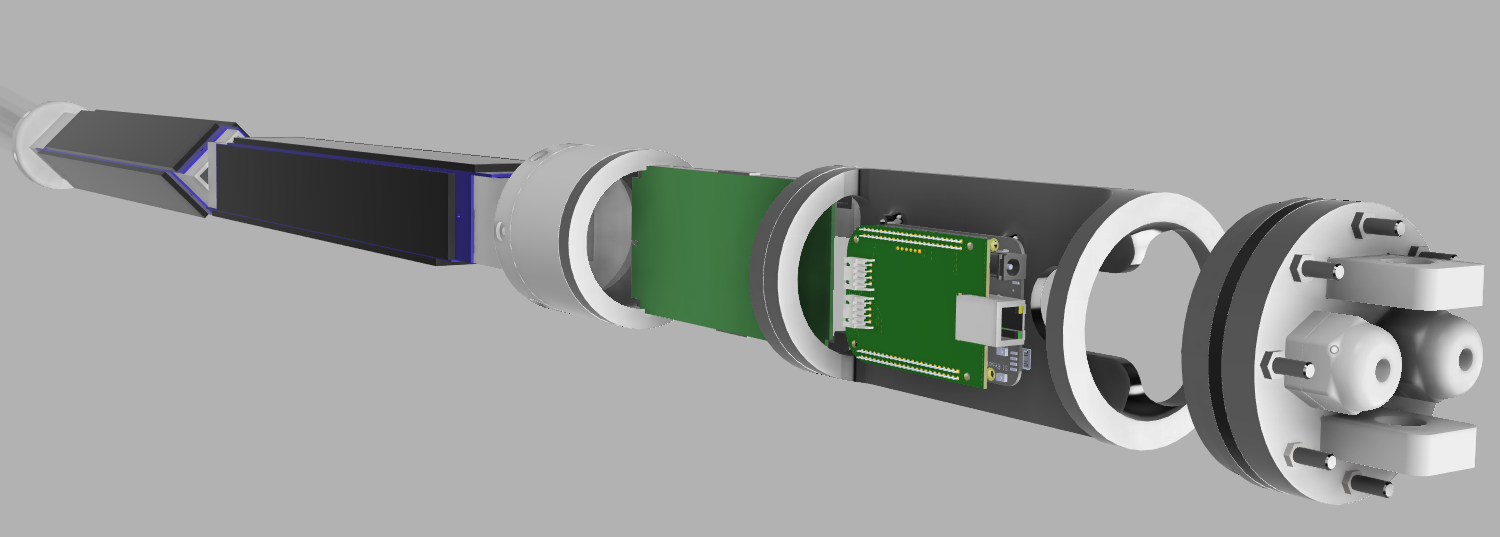}
\caption{Top: basic SiPM-on-tile module (left) and 32-channel half-detector unit (right). Bottom: 3D rendering of  64-channel unit (140 cm long, 80 mm in diameter) highlights the scintillators, DAQ module, Raspberry Pi, and end caps.}
\label{fig:SiPMunit}
\end{figure}

Figure~\ref{fig:SiPMunit} shows the SiPM board, an assembled 32-channel half-detector unit, and a 3D rendering of the 64-channel detector (140~cm long, 80~mm in diameter).

The 64-channel detector unit can in turn serve as a modular unit for longer detectors by stacking multiple modules, with each additional 64-channel unit rotated to further enhance azimuthal-angle reconstruction.

\FloatBarrier
\section{Simulation framework}
\label{sec:simulation}
The simulation was carried out using the DD4HEP framework~\cite{Frank:2014zya}, as interface to \textsc{Geant4}\cite{GEANT4:2002zbu} (v11.02.p2). We use the the FTFP\_BERT physics list with otherwise standard parameters. The input to the \textsc{Geant4} simulation consists of muons with trajectories generated according to the sea-level flux spectrum produced by \textsc{CRY}~\cite{Hagmann2007CRY}.

Simulated energy deposits were digitized according to the ADC capacity of the CAEN A5202, which is 13 bit. The production and transport of optical photons are not included in the \textsc{Geant4} simulation in order to reduce computational cost. Instead, we adopt a data-driven approach in which laboratory measurements are used to determine a smearing factor that is applied to the simulated energy depositions in an ad hoc manner.

\FloatBarrier
\section{Laboratory tests}
\label{sec:labtest}
The system was characterized with cosmic-ray measurements to evaluate the performance of a single SiPM-on-tile unit. The SiPM was biased at 43~V, corresponding to +5~V above its nominal breakdown voltage. An external trigger was provided to the CAEN A5202 by two scintillator tiles read out with 6~mm SENSL SiPMs, powered by their evaluation boards. The trigger tiles were positioned about 2.5~cm above and below the tile under test. A DRS4 digitizer generated the coincidence between the two trigger tiles and delivered the external trigger pulse (TTL) to the CAEN A5202. The readout of the tile under test was calibrated from arbitrary ADC units to the number of photoelectrons (PE) using the SiPM finger spectrum measured with a dedicated LED setup.   

\begin{figure}[h!]
    \centering
    \includegraphics[width=0.495\linewidth]{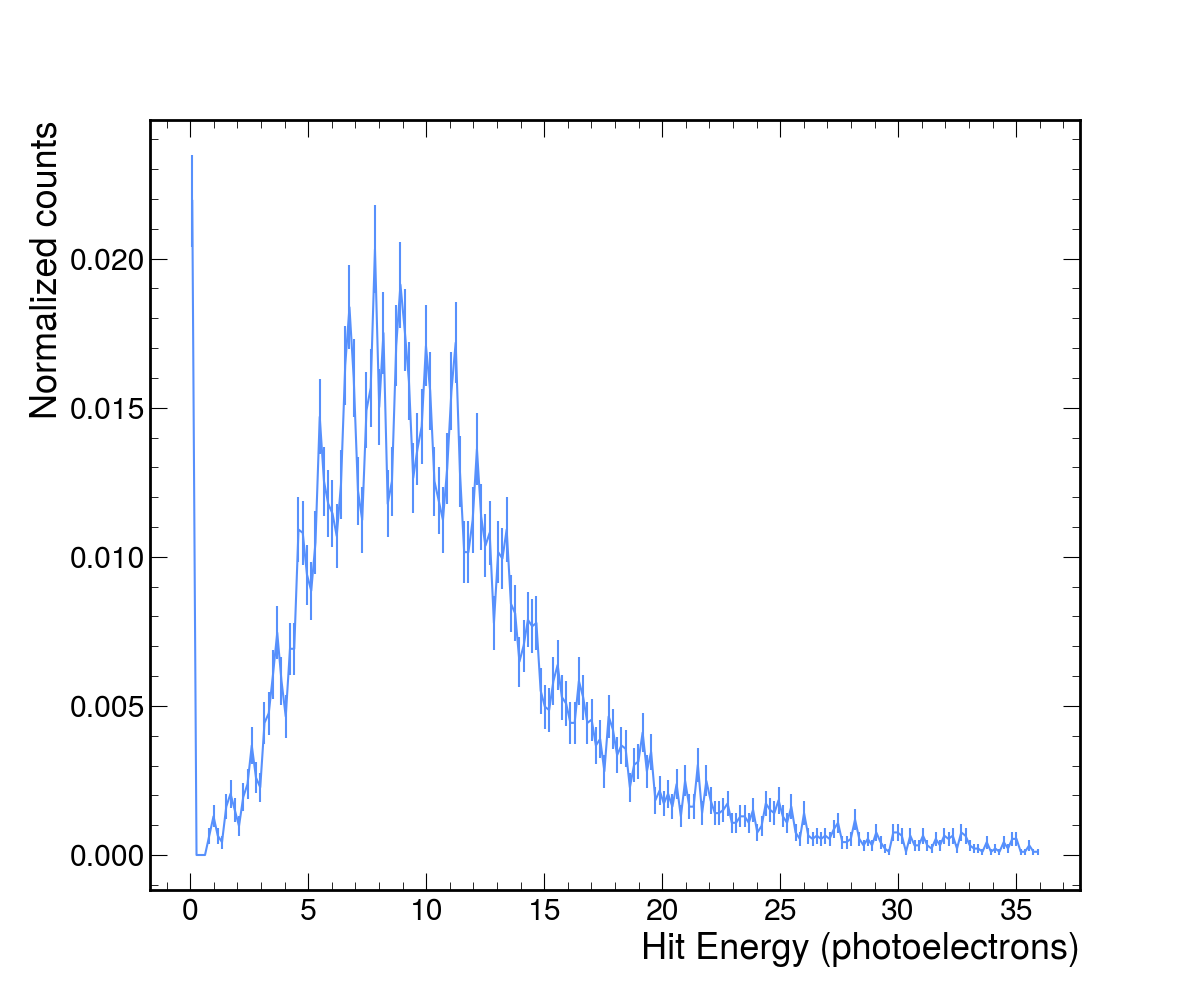}
    \includegraphics[width=0.495\linewidth]{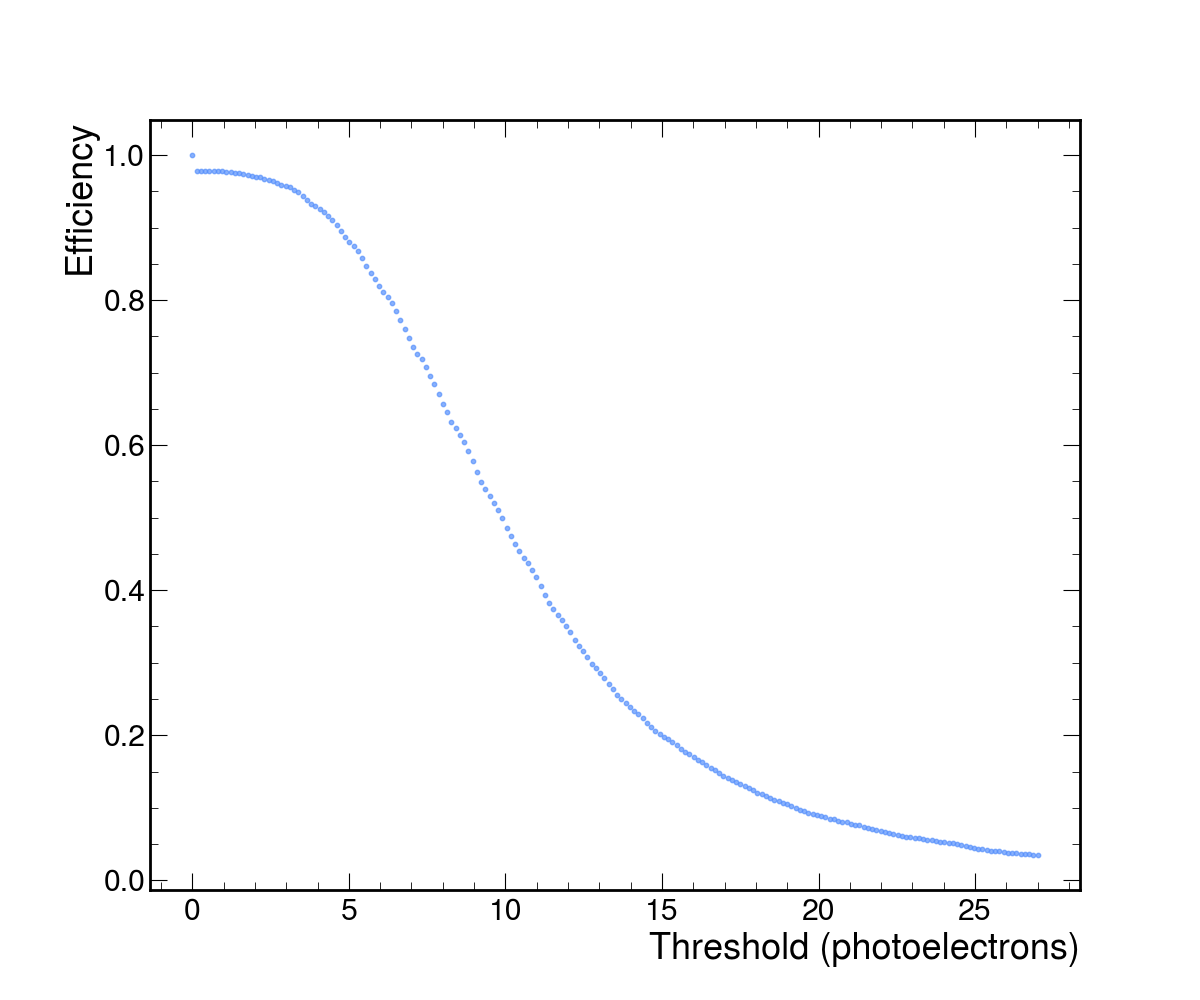}
    \caption{Light yield (left) and efficiency (right) measurement of single tile using cosmic-ray measurement with external trigger.}
    \label{fig:MIP_landau}
\end{figure}

Figure~\ref{fig:MIP_landau} shows the measured light yield. The observed distribution has a most probable value of about 8~PE. The distribution also exhibits individual photoelectron peaks that could be exploited for in situ calibration. From this data, the efficiency was evaluated for various thresholds and is shown in the right panel of Fig.~\ref{fig:MIP_landau}. The efficiency exceeds 96\% at a threshold of 3~PE. This threshold is sufficient to suppress the dark-count coincidence rate of two tiles to negligible levels. The intrinsic time resolution for this configuration is expected to be at the level of a few hundred picoseconds for a MIP deposition~\cite{Arratia:2023rdo}, thus in our design will be dominated by the resolution of the readout electronics which is about 0.5 ns.

\begin{figure}[h!]
    \centering
    \includegraphics[width=0.5\linewidth]{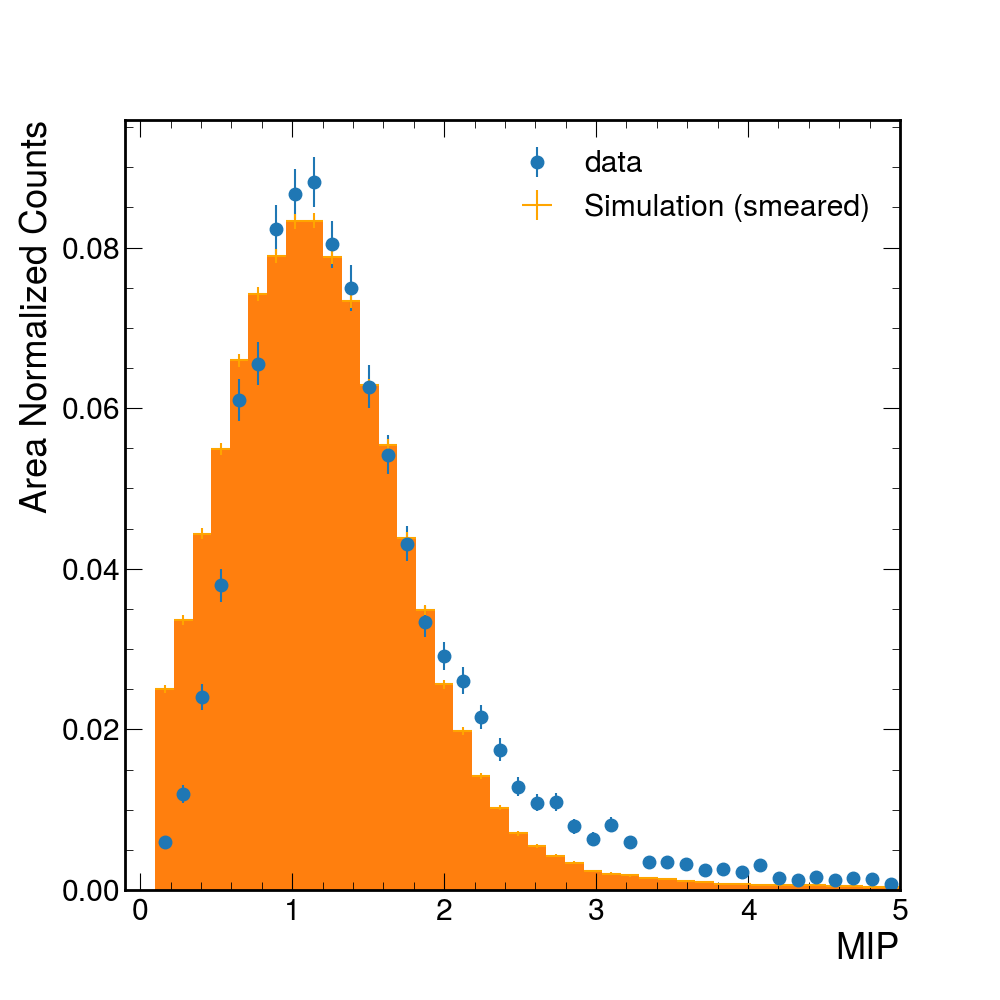}
    \caption{Landau measured in cosmic-ray setup of a single tile compared with simulations. A smearing factor was included in the simulation to roughly match the width of the peak in the data. }
    \label{fig:Landau_smeared}
\end{figure}

Figure~\ref{fig:Landau_smeared} shows the measured data (binned to average out the single-photon peaks) together with the simulation. The simulation reproduces the experimental setup, including the external trigger condition and geometry, and is smeared to approximately match the width of the Landau-like peak observed in data. Both data and simulation are calibrated such that the most probable value of the Landau-like distribution is unity, thereby defining the MIP scale. This MIP calibration can also be performed in situ on a channel-by-channel basis.

\FloatBarrier
\section{Simulated polar-angle performance}
\label{sec:performance}
The zenith angle $\theta_{\mathrm{reco}}$ and the azimuthal angle $\phi_{\mathrm{reco}}$ are reconstructed from the 3D hit positions according to

\begin{equation}
\phi_{\mathrm{reco}} = \arctan \left( \frac{V_y}{V_x} \right), 
\qquad
\theta_{\mathrm{reco}} =  \arctan \left( \frac{\left| V_z \right|}{\sqrt{(V_x)^{2}+(V_y)^{2}+(V_z)^{2}}} \right),
\end{equation}

where $V_x$, $V_y$, and $V_z$ denote the components of the reconstructed muon vector, obtained from a fit to the 3D hit positions. 

The zenith angle can also be inferred from the measured hit energies, since tracks closer to the zenith deposit, on average, more total energy than tracks at more oblique angles. The correlation between $\theta_{\mathrm{reco}}$ and the reconstructed energy is obtained from simulation, with a fit providing the most probable $\theta_{\mathrm{truth}}$ for a given reconstructed total energy $E_{\mathrm{reco}}$. While this standalone method performs worse than the 3D hit approach, it provides an independent estimate. 

A third approach uses a Graph Neural Network (GNN) that takes as input both the 3D hit positions and the measured energies to infer the muon track angle. Details of the GNN model can be found in Ref.~\cite{Milton:2024bqv}. 

\begin{figure}[h!]
    \centering
    \includegraphics[width=1.0\linewidth]{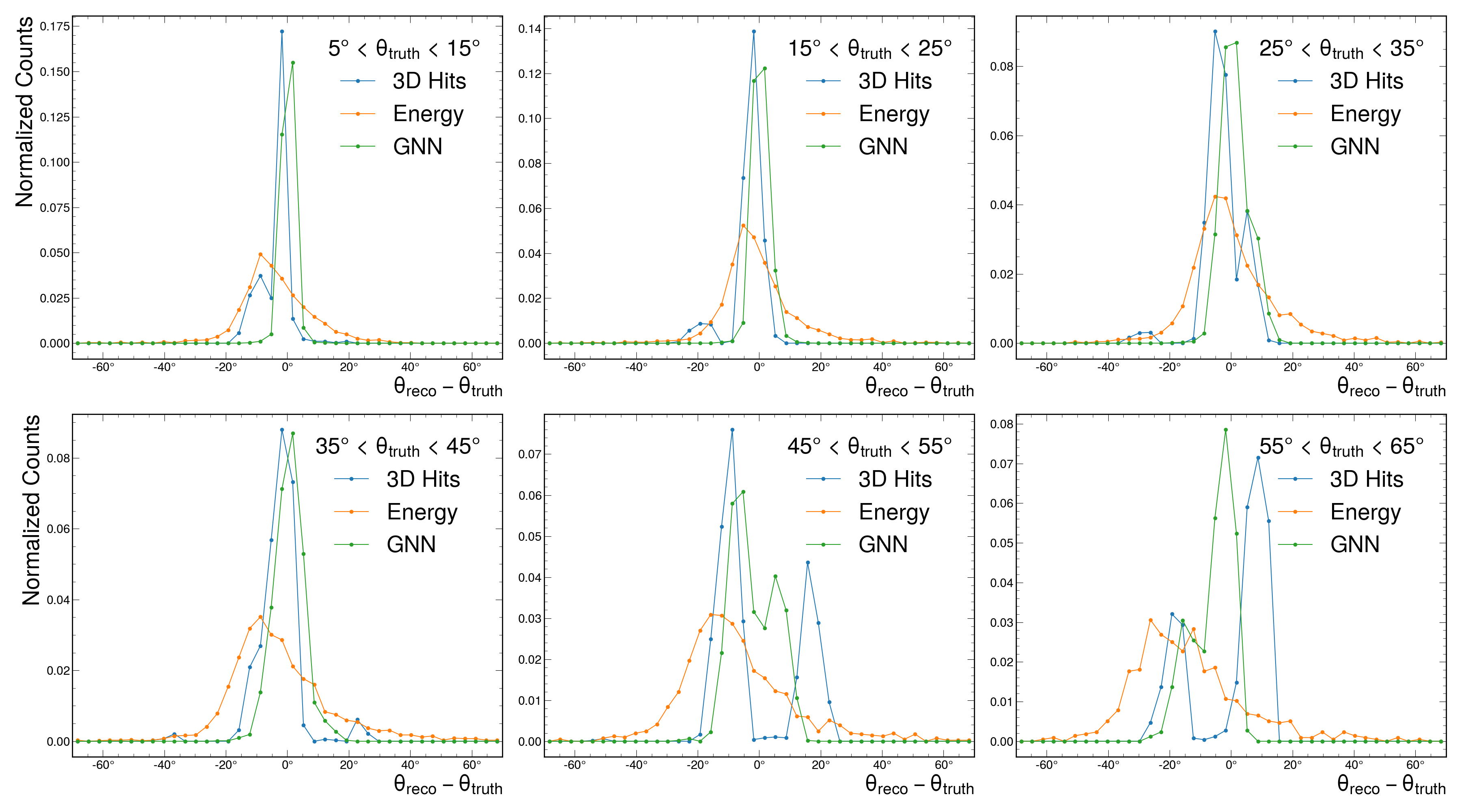}
    \caption{Zenith-angle performance in intervals of true zenith angle, obtained using the 3D hit method, the energy-based method, and the graph-neural network method.}
    \label{fig:polarperformance_bins}
\end{figure}

Figure~\ref{fig:polarperformance_bins} shows the zenith-angle performances of the 3D hit, energy-based, and GNN methods as a function of $\theta_{\mathrm{truth}}$, integrated over full azimuth. The 3D position method yields a Gaussian-like response up to $\theta_{\mathrm{truth}} \approx 45^\circ$, after which bimodal distributions appear, attributed to discretization effects due to the cell size. The energy-based method performs more poorly overall but yields a smoother response without discretization effects. The GNN achieves performance comparable to the 3D hit method, but with reduced bias, which is interpreted as the result of incorporating the additional energy information. Hit time information can also be incorporated in future studies. 

\begin{figure}[h!]
    \centering
    \includegraphics[width=0.495\linewidth]{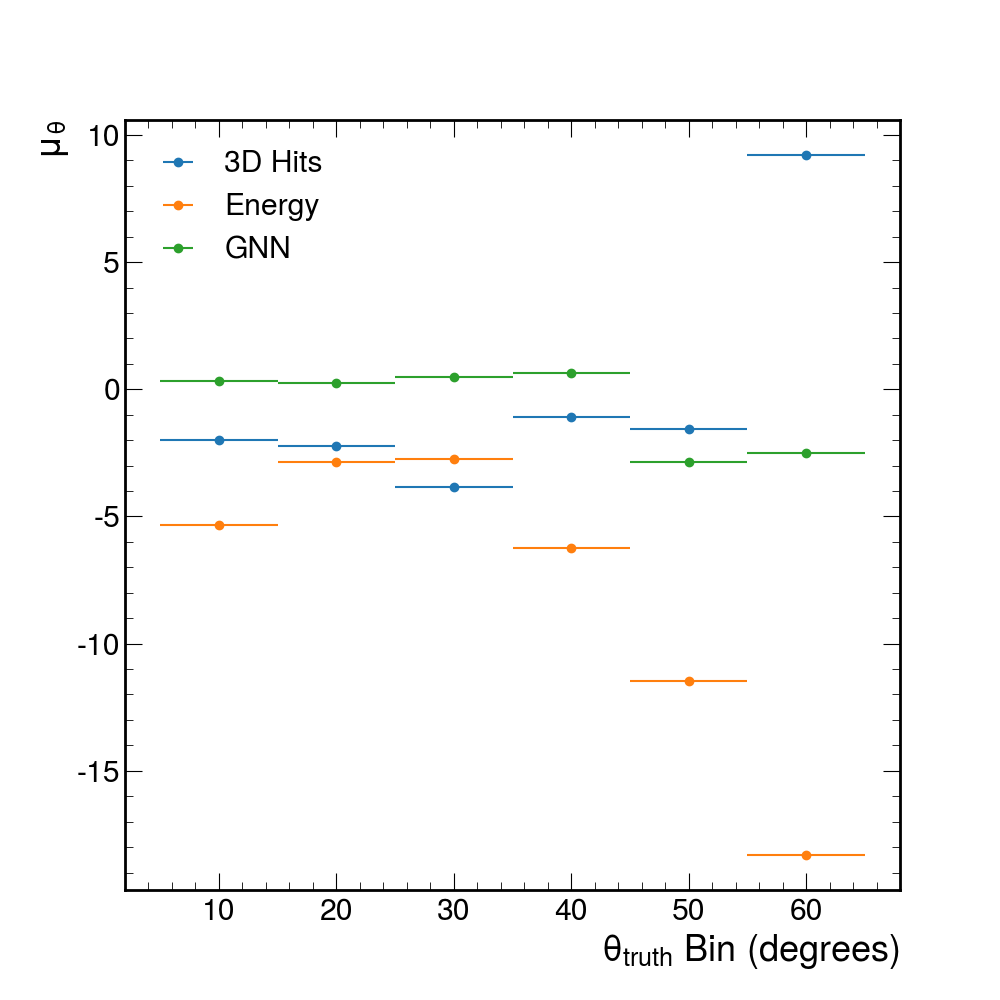}
        \includegraphics[width=0.495\linewidth]{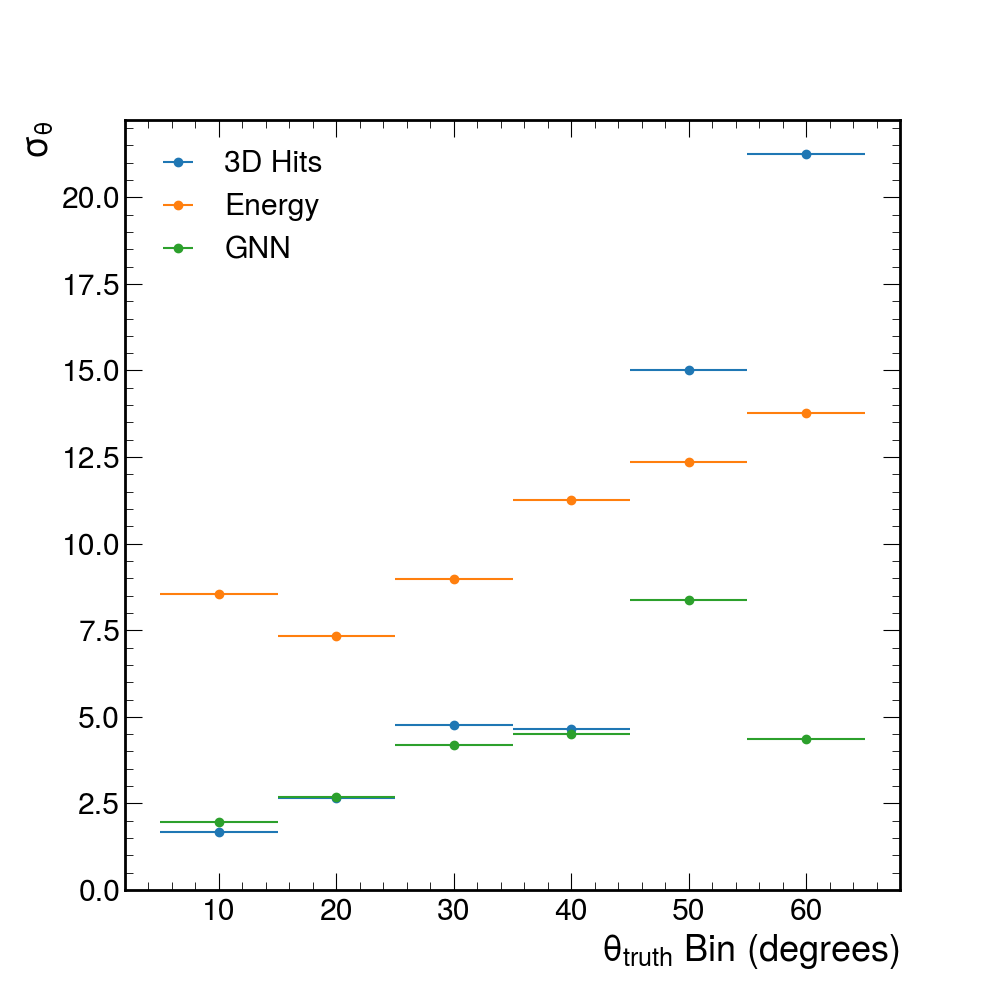}
    \caption{Zenith angle bias (left panel) and resolution (right panel) obtained using the 3D hit method, the energy-based method, and the graph-neural network method. }
    \label{fig:summaryperformance_polarangle}
\end{figure}

Figure~\ref{fig:summaryperformance_polarangle} summarizes the zenith-angle performance in terms of bias and resolution obtained from Gaussian fits to the distributions shown in Fig.~\ref{fig:polarperformance_bins}. These Gaussian metrics are presented for illustration, although they do not always capture the underlying distributions, which can exhibit multimodality due to discretization effects and ambiguities in zenith reconstruction. Solutions to these issues will be investigated in future designs, which may also incorporate timing information.

\FloatBarrier
\section{Conclusions}
\label{sec:conclusions}
We have presented the first borehole muon detector based on a scintillator tile design. The modular detector fits within standard HQ boreholes, measures about 140~cm in length and 80~mm in diameter, and contains 64 channels. It achieves a zenith resolution of 1.5--4.0$^\circ$, depending on the zenith angle. Although the azimuthal granularity is coarser, it can be improved by combining measurements from multiple detectors rotated axially with respect to each other. By enabling three-dimensional hit reconstruction with a single scintillator layer, the SiPM-on-tile concept reduces detector diameter, eliminates the need for picosecond timing electronics, and lowers production costs through the use of smaller 1.3~mm SiPMs. Overall, this detector provides a compact, rugged, and cost-effective solution for muon tomography.

\acknowledgments

We acknowledge the support from a UC Riverside Academic Senate Research Grant and ANID PIA/APOYO AFB230003. We thank Ryan Milton for his help with the GNN.  
\bibliographystyle{JHEP}

\bibliography{biblio} 

\providecommand{\href}[2]{#2}\begingroup\raggedright\begin{thebibliography}{10}

\bibitem{Tanaka2023Muography}
H.K.M.~Tanaka, C.~Bozza, A.~Bross, E.~Cantoni, O.~Catalano, G.~Cerretto et~al., \emph{Muography}, \href{https://doi.org/10.1038/s43586-023-00270-7}{\emph{Nature Reviews Methods Primers} {\bfseries 3} (2023) 88}.

\bibitem{IAEA2022MuonImaging}
\emph{Muon Imaging}, no.~2012 in IAEA TECDOC Series, International Atomic Energy Agency, Vienna (2022).

\bibitem{Levy1985BoreholeMuon}
R.H.~Levy and P.M.~Mockett, ``Method and apparatus for determining the density characteristics of underground earth formations.'' U.S. Patent US4504438A, Mar., 1985.

\bibitem{Schwitters2004MuonTomography}
R.F.~Schwitters, \emph{A detector for muon tomography},  Conceptual Design Report \href{https://www.hep.utexas.edu/mayamuon/files/papers/cdr.pdf}{https://www.hep.utexas.edu/mayamuon/files/papers/cdr.pdf}, University of Texas at Austin, High Energy Physics Group (UT MAYA Muon Collaboration) (June, 2004).

\bibitem{fermi2007university}
F.N.A.~Laboratory, U.S.D.~of~Energy. Office~of Scientific and T.~Information, \emph{The University of Texas Maya Muon Project}, United States. Department of Energy. Office of Science (2007).

\bibitem{SchreinerIII:2016ezv}
H.F.~Schreiner~III, \emph{{Methods and Simulations of Muon Tomography and Reconstruction}}, Ph.D. thesis, Texas U., 2016.

\bibitem{BASSET2006298}
M.~Basset, S.~Ansoldi, M.~Bari, R.~Battiston, S.~Blasko, F.~Coren et~al., \emph{Mgr: An innovative, low-cost and compact cosmic-ray detector}, \href{https://doi.org/10.1016/j.nima.2006.05.099}{\emph{Nuclear Instruments and Methods in Physics Research Section A: Accelerators, Spectrometers, Detectors and Associated Equipment} {\bfseries 567} (2006) 298}.

\bibitem{Klinger:2015gva}
J.~Klinger et~al., \emph{{Simulation of muon radiography for monitoring CO2 stored in a geological reservoir}}, \href{https://doi.org/10.1016/j.ijggc.2015.09.010}{\emph{International Journal of Greenhouse Gas Control} {\bfseries 42} (2015) 644} [\href{https://arxiv.org/abs/1510.06372}{{\ttfamily 1510.06372}}].

\bibitem{Gluyas2018CO2Muon}
J.~Gluyas, L.~Thompson, D.~Allen, C.~Benton, P.~Chadwick, S.~Clark et~al., \emph{Passive, continuous monitoring of carbon dioxide geostorage using muon tomography}, \href{https://doi.org/10.1098/rsta.2018.0059}{\emph{Philosophical Transactions of the Royal Society A: Mathematical, Physical and Engineering Sciences} {\bfseries 377} (2019) 20180059}.

\bibitem{Yaish2017GasMuon}
D.~Yaish, A.~Harel and Y.~Kolkovich, ``Systems and methods for underground exploration using cosmic rays muons.'' U.S. Patent US10585208B1, Mar., 2017.

\bibitem{Harel2019Lingacom}
A.~Harel and D.~Yaish, \emph{Lingacom muography}, \href{https://doi.org/10.1098/rsta.2018.0133}{\emph{Philosophical Transactions of the Royal Society A: Mathematical, Physical and Engineering Sciences} {\bfseries 377} (2019) 20180133}.

\bibitem{Bonneville2017MuonDetector}
A.~Bonneville, R.T.~Kouzes, J.~Yamaoka et~al., \emph{A novel muon detector for borehole density tomography}, \href{https://doi.org/10.1016/j.nima.2017.01.023}{\emph{Nuclear Instruments and Methods in Physics Research Section A: Accelerators, Spectrometers, Detectors and Associated Equipment} {\bfseries 851} (2017) 108}.

\bibitem{Flygare2018MuonBorehole}
J.D.~Flygare, A.~Bonneville, R.T.~Kouzes, J.~Yamaoka and A.~Lintereur, \emph{Muon borehole detector design for use in 4-d density overburden monitoring}, \href{https://doi.org/10.1109/TNS.2018.2869196}{\emph{IEEE Transactions on Nuclear Science} {\bfseries 65} (2018) 2724}.

\bibitem{Olah2018MWPCBorehole}
L.~Oláh, G.~Hamar, S.~Miyamoto, H.K.M.~Tanaka and D.~Varga, \emph{The first prototype of an mwpc-based borehole detector and its application for muography of an underground pillar}, \href{https://doi.org/10.3124/segj.71.161}{\emph{BUTSURI-TANSA (Geophysical Exploration)} {\bfseries 71} (2018) 161}.

\bibitem{Hamar2022Portable}
G.~Hamar, G.~Surányi, D.~Varga et~al., \emph{Underground muography with portable gaseous detectors},  in \emph{Journal of Physics: Conference Series}, vol.~2374, p.~012186, IOP Publishing, 2022, \href{https://doi.org/10.1088/1742-6596/2374/1/012186}{DOI}.

\bibitem{HiroshiSuenaga2018}
H.~Suenaga and H.~Tanaka, \emph{Development of a muon density logging tool for use in boreholes and field applications}, \href{https://doi.org/10.3124/segj.71.148}{\emph{BUTSURI-TANSA(Geophysical Exploration)} {\bfseries 71} (2018) 148}.

\bibitem{Schouten:2018pqz}
D.~Schouten and P.~Ledru, \emph{{Muon Tomography Applied to a Dense Uranium Deposit at the McArthur River Mine}}, \href{https://doi.org/10.1029/2018jb015626}{\emph{J. Geophys. Res. Solid Earth} {\bfseries 123} (2018) 8637}.

\bibitem{Schouten2024Helical}
D.W.~Schouten and J.~van Nieuwkoop, ``Apparatus, system and method regarding borehole muon detector for muon radiography and tomography.'' U.S. Patent US11994645B2, assigned to Ideon Technologies Inc., May, 2024.

\bibitem{Holma2020_US11892463B2}
M.~Holma and P.~Kuusiniemi, ``System and method for material density distribution survey based on cosmic muon detection.'' U.S. Patent US11892463B2, Jan., 2024.

\bibitem{Schouten_drift}
D.W.~Schouten, ``Drift tube borehole muon detector system, apparatus, and method for muon radiography and tomography.'' U.S. Patent US12007527B2, assigned to Ideon Technologies Inc., June, 2024.

\bibitem{Cimmino2021BoreholeDetector}
L.~Cimmino, F.~Ambrosino, A.~Anastasio et~al., \emph{A new cylindrical borehole detector for radiographic imaging with muons}, \href{https://doi.org/10.1038/s41598-021-96247-1}{\emph{Scientific Reports} {\bfseries 11} (2021) 17425}.

\bibitem{Ambrosino2022Borehole}
F.~Ambrosino, L.~Cimmino, A.~Anastasio et~al., \emph{A cylindrical muon detector for borehole applications}, \href{https://doi.org/10.3390/instruments6040064}{\emph{Instruments} {\bfseries 6} (2022) 64}.

\bibitem{Saracino2023Borehole}
G.~Saracino, F.~Ambrosino, A.~Anastasio et~al., \emph{A new cylindrical detector for borehole muon radiography}, \href{https://doi.org/10.1016/j.nima.2022.167995}{\emph{Nuclear Instruments and Methods in Physics Research Section A: Accelerators, Spectrometers, Detectors and Associated Equipment} {\bfseries 1048} (2023) 167995}.

\bibitem{Gauvreau2024B2}
J.~Gauvreau, D.~Snowden{-}Ifft and J.~Kirkpatrick, ``Muon detector for muon tomography.'' U.S. Patent US12105230B2, assigned to KoBold Metals Company, Oct., 2024.

\bibitem{Yamazaki2022}
K.~Yamazaki, A.~Taketa, D.~Ikeda et~al., \emph{Development of detector and method for density structure measurement of fault zones using cosmic ray muons}, \href{https://doi.org/10.1016/j.nima.2022.166518}{\emph{Nuclear Instruments and Methods in Physics Research Section A: Accelerators, Spectrometers, Detectors and Associated Equipment} {\bfseries 1031} (2022) 166518}.

\bibitem{Yao:2023oyn}
K.~Yao et~al., \emph{{Concept design and feasibility study of novel calorimeter-type borehole muon detector}}, \href{https://doi.org/10.1016/j.nima.2023.168074}{\emph{Nucl. Instrum. Meth. A} {\bfseries 1049} (2023) 168074}.

\bibitem{Li2024Omnidirectional}
Z.~Li, J.~Zhang, W.~Liu et~al., \emph{Omnidirectional borehole detector for muography: Design and performance evaluation}, \href{https://doi.org/10.1016/j.nima.2024.169568}{\emph{Nuclear Instruments and Methods in Physics Research Section A: Accelerators, Spectrometers, Detectors and Associated Equipment} {\bfseries 1065} (2024) 169568}.

\bibitem{Zhao_2024}
Y.~Zhao and Y.~Guan, \emph{The cosmic ray muons tomography system with wavelength shifting fiber and plastic scintillator plane and vertical borehole detectors and improved 3d image reconstruction algorithm: a simulation study}, \href{https://doi.org/10.1088/1748-0221/19/11/T11006}{\emph{Journal of Instrumentation} {\bfseries 19} (2024) T11006}.

\bibitem{Yin2024Borehole}
L.~Yin, S.~Luo, G.~Zeng, K.~Wang, S.~Feng, S.Y.~Feng et~al., \emph{Simulation research of a borehole muon detector for deep vein exploration}, \href{https://doi.org/10.1088/1748-0221/19/11/P11027}{\emph{Journal of Instrumentation} {\bfseries 19} (2024) P11027}.

\bibitem{Sefkow:2015hna}
F.~Sefkow, A.~White, K.~Kawagoe, R.~P\"oschl and J.~Repond, \emph{{Experimental Tests of Particle Flow Calorimetry}}, \href{https://doi.org/10.1103/RevModPhys.88.015003}{\emph{Rev. Mod. Phys.} {\bfseries 88} (2016) 015003} [\href{https://arxiv.org/abs/1507.05893}{{\ttfamily 1507.05893}}].

\bibitem{CALICE:2022uwn}
{\scshape CALICE} collaboration, \emph{{Design, construction and commissioning of a technological prototype of a highly granular SiPM-on-tile scintillator-steel hadronic calorimeter}}, \href{https://doi.org/10.1088/1748-0221/18/11/P11018}{\emph{JINST} {\bfseries 18} (2023) P11018} [\href{https://arxiv.org/abs/2209.15327}{{\ttfamily 2209.15327}}].

\bibitem{CMS:2022jvd}
{\scshape CMS, CALICE} collaboration, \emph{{Performance of the CMS High Granularity Calorimeter prototype to charged pion beams of 20{\textendash}300 GeV/c}}, \href{https://doi.org/10.1088/1748-0221/18/08/P08014}{\emph{JINST} {\bfseries 18} (2023) P08014} [\href{https://arxiv.org/abs/2211.04740}{{\ttfamily 2211.04740}}].

\bibitem{Belloni_2021}
A.~Belloni, Y.~Chen, A.~Dyshkant, T.~Edberg, S.~Eno, J.~Freeman et~al., \emph{Test beam study of {SiPM}-on-tile configurations}, \href{https://doi.org/10.1088/1748-0221/16/07/p07022}{\emph{Journal of Instrumentation} {\bfseries 16} (2021) P07022}.

\bibitem{Frank:2014zya}
M.~Frank, F.~Gaede, C.~Grefe and P.~Mato, \emph{{DD4hep: A Detector Description Toolkit for High Energy Physics Experiments}}, \href{https://doi.org/10.1088/1742-6596/513/2/022010}{\emph{J. Phys. Conf. Ser.} {\bfseries 513} (2014) 022010}.

\bibitem{GEANT4:2002zbu}
{\scshape GEANT4} collaboration, \emph{{GEANT4--a simulation toolkit}}, \href{https://doi.org/10.1016/S0168-9002(03)01368-8}{\emph{Nucl. Instrum. Meth. A} {\bfseries 506} (2003) 250}.

\bibitem{Hagmann2007CRY}
C.~Hagmann, D.~Lange and D.~Wright, \emph{Cosmic‐ray shower generator (cry) for monte carlo transport codes},  in \emph{2007 IEEE Nuclear Science Symposium Conference Record (NSS-MIC)}, vol.~2, pp.~1143--1146, 2007, \href{https://doi.org/10.1109/NSSMIC.2007.4437209}{DOI}.

\bibitem{Arratia:2023rdo}
M.~Arratia, L.~Garabito~Ruiz, J.~Huang, S.J.~Paul, S.~Preins and M.~Rodriguez, \emph{{Studies of time resolution, light yield, and crosstalk using SiPM-on-tile calorimetry for the future Electron-Ion Collider}}, \href{https://doi.org/10.1088/1748-0221/18/05/P05045}{\emph{JINST} {\bfseries 18} (2023) P05045} [\href{https://arxiv.org/abs/2302.03646}{{\ttfamily 2302.03646}}].

\bibitem{Milton:2024bqv}
R.~Milton, S.J.~Paul, B.~Schmookler, M.~Arratia, P.~Karande, A.~Angerami et~al., \emph{{Design and simulation of a SiPM-on-tile ZDC for the future EIC, and its performance with graph neural networks}}, \href{https://doi.org/10.1016/j.nima.2025.170613}{\emph{Nucl. Instrum. Meth. A} {\bfseries 1079} (2025) 170613} [\href{https://arxiv.org/abs/2406.12877}{{\ttfamily 2406.12877}}].

\end{thebibliography}\endgroup

\end{document}